\documentclass[pra,bibnotes,twocolumn]{revtex4}
\usepackage{graphicx}
\usepackage{hyperref}
\usepackage{graphicx}
\usepackage{amsmath,amssymb}
\usepackage{float}
\usepackage{bm}

\begin{document}
\draft
\def\ds{\displaystyle}
\title{ Nonlinear non-Hermitian skin effect}
\author{Cem Yuce}
\email{cyuce@eskisehir.edu.tr}
\address{Department of Physics, Faculty of Science, Eskisehir Technical University, Turkey }

\date{\today}
\begin{abstract}
Distant boundaries in linear non-Hermitian lattices can dramatically change energy eigenvalues and corresponding eigenstates in a nonlocal way. This effect is known as non-Hermitian skin effect (NHSE). Combining non-Hermitian skin effect with nonlinear effects can give rise to a host of novel phenomenas, which may be used for nonlinear structure designs. Here we study nonlinear non-Hermitian skin effect and explore nonlocal and substantial effects of edges on stationary nonlinear solutions. We show that fractal and continuum bands arise in a long lattice governed by a nonreciprocal discrete nonlinear Schrodinger equation. We show that stationary solutions are localized at the edge in the continuum band. We consider a non-Hermitian Ablowitz-Ladik model and show that nonlinear exceptional point disappears if the lattice is infinitely long.
\end{abstract}
\maketitle

\section{Introduction}

It has recently been predicted \cite{nhs1,nhs2,nhs3,nhs4,nhs5,nhs6,nhs7,nhs8,nhs9,cemAnnals,nhs10,nhs12,nhs13,nhs14,nhs15,nhs16,nhs17,cluster} and experimentally realized \cite{nhs19,nhs20,nhs21,funnel,nhs18} that the transition from periodic boundary conditions (PBC) to open boundary conditions (OBC) in sufficiently long non-reciprocal lattices can become substantially non-perturbative even though the corresponding non-Hermitian Hamiltonians are perturbatively different from each other. More specifically, the spectra and corresponding eigenstates with or without boundaries can be significantly different in non-Hermitian systems. Such a non-local change due to the distant boundaries is unique to non-Hermitian systems and this effect is known as non-Hermitian skin effect. It has an interesting consequences of strong localizations of eigenstates at one edge, which in turn leads to so called funneling effect \cite{funnel} since any form of initial wave packet always moves to the edge without backscattering from the edge. Furthermore, the bulk-boundary correspondence fails due to such a non-local change of the eigenstates. \\
NHSE has been so far investigated only for linear non-reciprocal systems \cite{refek4,refek5,refek6,refek7,refek8,refek9,refek10,refek11,refek12,refek13,refek14,CYHR,ourpaperwx}. It is an open question whether NHSE occurs in nonlinear non-Hermitian systems. In this paper, we study this problem and explore NHSE for some nonlinear models. The nonlinearity renders the problem much more difficult since the total number of stationary solutions increases exponentially with lattice size as opposed to the linear systems where the total number of eigenstates increases linearly with total number of lattice sites. Furthermore, analytical solutions are rare in nonlinear systems. Complete characterization of nonlinear problems can be possible for a simple system such as nonlinear dimer and trimer \cite{dimer1,dimer2,dimer3} but NHSE should be studied for a long lattice to understand the nonlocal effect of distant boundaries. Unfortunately, the complexity of the problem exponentially increases as the nonlinear lattice size is increased. Therefore, it is challenging to show whether localizations of stationary solutions in nonlinear domain occurs as a result of distant boundaries. Proving that spectra for PBC and OBC change significantly due to the presence boundaries is also challenging since nonlinear system are generally chaotic. In this paper, we consider a nonreciprocal discrete nonlinear Schrodinger equation and a non-Hermitian Ablowitz-Ladik equation. We show that fractal and  continuum bands occur for the latter system. We perform some analytical solutions and show that all stationary solutions are localized in the continuum band. We predict that infinitely and sufficiently long non-Hermitian lattices can have drastically different stationary solutions.

\section{ Nonreciprocal discrete nonlinear Schrodinger equation}

We start with the 1D nonreciprocal discrete nonlinear Schrodinger equation for the complex field amplitude $\ds{  \psi_j }$ at site $\ds{j}$
\begin{equation}\label{rof64oalk2cy} 
\psi_{j+1}+\gamma ~\psi_{j-1}+g~ |\psi_{j}  |^2~ \psi_{j} =\omega~ \psi_{j} 
\end{equation} 
where $\ds{ j=0,1,2,...,N }$ and $\ds{N+1}$ is the total number of lattice sites, $\ds{ g>0   }$ is the positive nonlinear interaction strength, $\ds{ 0\leq \gamma<1 }$ is the non-Hermitian degree and $\ds{  \omega }$ is the frequency.\\
Consider first a long linear lattice, $\ds{g=0}$. In this case, the dramatic role of the boundary conditions on the entire spectrum can easily be understood at $\ds{ \gamma=0 }$. For PBC, all eigenstates are extended ($\ds{ \psi_j=  \psi_0 ~e^{ik j} }$ corresponding to complex eigenvalues $\ds {e^{ik} }$, where $\ds{\psi_0}$ is the complex field amplitude at the left edge). On the other hand, for OBC, an exceptional point occurs and all eigenstates coalesce to a zero-energy eigenstate localized at the left edge. In fact, the OBC and PBC lattices become perturbatively different from each other when $\ds{N}$ is sufficiently large. However, such a local perturbative difference has nonlocal and nonperturbative effects on the  the spectrum and the eigenstates. This is the essence of NHSE. Suppose next that $\ds{\gamma}$ is non-zero but a small number. In this case, all eigenstates are still extended for PBC and the corresponding eigenvalues, $\ds {e^{ik}+\gamma ~e^{-ik} }$, change perturbatively with $\ds{\gamma}$. In the case of OBC, the exceptional eigenstate for $\gamma=0$ is split into $\ds{N+1}$ distinct eigenstates, which exhibit strong localization at the left edge. Based on the above discussions, we are tempted to begin to study nonlinear NHSE at $\ds{\gamma=0}$ and then make a generalization to small values of $\ds{\gamma}$. Below, we follow this approach.\\
Consider next a long nonlinear lattice, $\ds{g\neq0}$. The plane waves $\ds{  \psi_{j}=\psi_0~ e^{ikj}  }$ constitute a family of solutions for PBC with complex valued frequencies $\ds{     \omega_{k}=e^{ik  }+{\gamma}~e^{-ik  }+g| \psi_0|^2   }$. This picture changes drastically if we apply OBC, $\ds{ \psi_{-1} =\psi_{N+1}=0 }$. Let us look for all family of the stationary solutions for OBC. Fortunately, we get some analytical solutions at $\ds{  \gamma=0 }$. In this case, we get the nonlinear recurrence relation $\ds{  \psi_{j+1} = \left(  \omega-g|\psi_j|^2   \right)  \psi_j  }$, which can be solved recursively by setting the initial value, $\ds{  \psi_{0} }$, at the left edge and obtaining each successive term of the sequence from the preceding terms. Note that if a term in the sequence becomes zero, then all other successive terms are also zero. This implies that the open boundary condition at the right edge is automatically satisfied. This is possible only when the frequency $\ds{\omega}$ takes some certain values at fixed $\ds{|\psi_0|}$. The first such solution is given by
\begin{equation}\label{ddekcy23uds2} 
 \psi_j=\psi_0~ \delta_{j,0} ~,~~~~\omega_0=g| \psi_0|^2
\end{equation}
The next solution can be found by assuming that only the first two sites on the left are populated. There are two such solutions
\begin{eqnarray}\label{cs3aek2} 
 \psi_j&=&\psi_0~ \delta_{j,0} +  (\omega_{\mp}-g| \psi_0|^2) ~ \psi_0~    \delta_{j,1}       \nonumber \\
\omega_{\mp}&=&  g |\psi_0|^2+  \frac{1\mp\sqrt{1+  4g^2 |\psi_0|^4}    }{2 g  |\psi_0|^2 } 
\end{eqnarray}
where $\ds{\omega_{+}}$ is always greater than $\ds{2.6}$, while $\ds{\omega_{-}}$ can take values from zero to infinity. \\ 
It is cumbersome to write all $\ds{6}$ solutions when only the first three sites are occupied, at which complex frequency values appear. 
In this case, the frequencies should be found numerically by solving a sextic polynomial equation. One can go in this way and find the spectrum numerically for a given value of $N$. Unfortunately, one needs high computational power as the total number of solutions, $\ds{3^{N-1}}$, grows exponentially with the lattice size. Note that NHSE is more visible for a long lattice and this poses difficulties to study this effect in nonlinear domain. \\
$\ds{ \it{Semi-infinite~ lattice :}  }$ Suppose that $\ds{N \rightarrow \infty}$ and $\ds{\gamma=0}$. In this case, solving our problem as outlined above is not possible. One may instead start with arbitrary values of $\ds{ \psi_0 }$ and $\ds{\omega}$ and then obtain the terms of the infinite sequence repeatedly in a self-similar way. Unfortunately, this trial and error method does not work well as our system is highly chaotic. In other words, an arbitrarily small perturbation to $\ds{ \psi_0 }$ such as roundoff errors at fixed $\ds{\omega}$ (or vice versa) can produce tremendous changes in the sequence. As a numerical illustration, we iteratively solve Eq. (\ref{rof64oalk2cy}) when $\ds{\psi_0=2g=2}$ and $\ds{ \omega=\omega_{+}=4+  \frac{1+\sqrt{  65  }    }{8  }   } $ at which we are supposed to obtain the solution (\ref{cs3aek2}) numerically (we evaluate $\ds{\omega_+}$ numerically to 21-digit precision to make the roundoff error very small). We see that the field amplitudes become almost zero within a few sites from the left edge but the iteration diverges to infinity when we keep iterating further (for example, $\ds{|\psi_{40}| \approx 10^{90} }$). This shows that the number of numerical precision must be taken sufficiently large to avoid divergence in a finite long lattice. For the semi-infinite lattice, roundoff errors are serious problems. A question arises. For what $\ds{\omega}$ and $\ds{ \psi_0 }$ will the field amplitudes remain bounded and eventually converge to zero? Below, we show that $\ds{\omega}$ vs $\ds{| \psi_0|^2 }$ exhibits $\it{fractal}$ and $\it{continuum}$ structures.  \\
Let us obtain frequency distributions for the semi-infinite lattice. We have already obtained three frequencies exactly in Eqs. (\ref{ddekcy23uds2}, \ref{cs3aek2}). Note that the solutions with $\ds{\omega_{+}}$ ($\ds{\omega_{-}}$) are unstable for all (large) values of $\ds{|\psi_0|}$. To see this, we introduce a very small perturbation at $j=1$ site i. e., $\psi_j\rightarrow\psi_j+\delta\psi_1~\delta_{j,1}$, where $ |\delta\psi_1|<<|\psi_1|   $. Then we obtain $\ds{   \psi_2\approx-2g\psi_1^2 \delta\psi_1 }$, $\ds{   \psi_3\approx \psi_2 ~ \omega_{\mp}     }$, $\ds{   \psi_4\approx    \psi_2   ~ \omega_{\mp}^2 }$ and so on (for real valued field amplitudes). We see that the perturbation grows at each step and diverges at infinity if $\ds{|  \omega_{\mp}  |>1}$. Let us now focus on the frequency $\ds{\omega_{+}}$ and look for some other solutions with frequencies close to $\ds{\omega_{+}}$, i. e., $\ds{ \omega= \omega_{+}+\delta \omega_{+}}$ where $\ds{ |\delta \omega_{+} |<<   \omega_{+}}$. To find such stationary solutions for a finite but long lattice, we make iterations by slightly varying $\ds{ \psi_0}$ and $\ds{\omega}$ until it's clear that the right open boundary condition is satisfied. We numerically find that there exists so many such values of $\ds{ \delta \omega_{+} }$. This leads to $\it{ fractal ~band}$ structure around $\ds{\omega_{+}}$. The fractal band can be thought of as the collection of the individual frequency levels that are very close but not continuous and repeat themselves in a self-similar way. This band is non-differentiable and has unstable solutions. To get a visual understanding of this irregular geometric structure, we plot such a fractal band for real valued frequencies when there are $7$ lattice sites in Fig.1. We see that the frequencies are piled up around $\ds{\omega_{\mp}}$ and also $\ds{\omega_0}$ when $\ds{\psi_0>2}$. The feature of self-similarity can be understood if we zoom in/out the fractal band. As an example, we zoom in the small area depicted in the red rectangle in the bottom left corner of (a) and plot it in (b). As you see, the two plots look similar, which is a common feature of fractal structures. The fractal band is generally complex and its imaginary part can also be studied in a similar way. \\
\begin{figure}[t]\label{2678ik0}
\includegraphics[width=4.1cm]{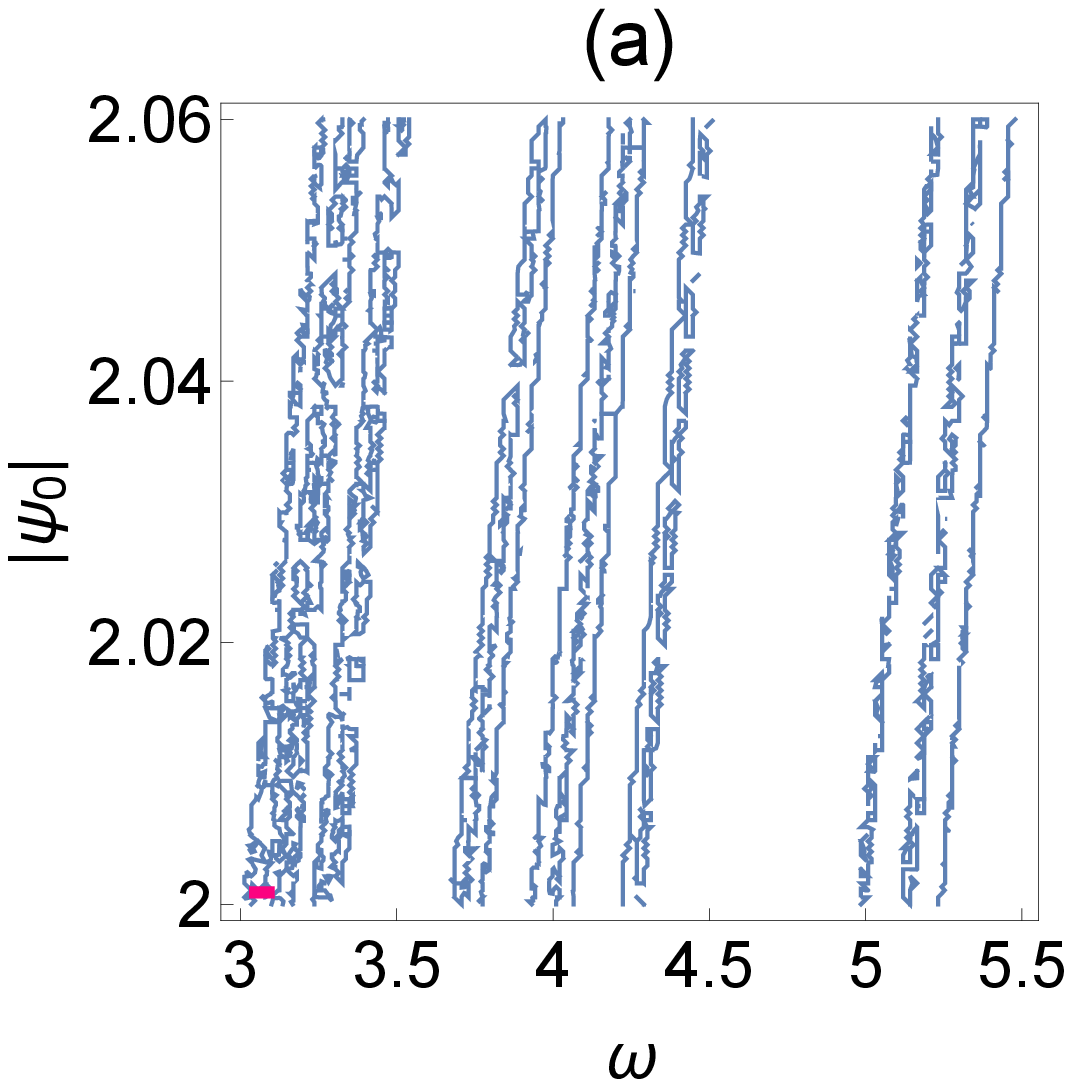}
\includegraphics[width=4.4cm]{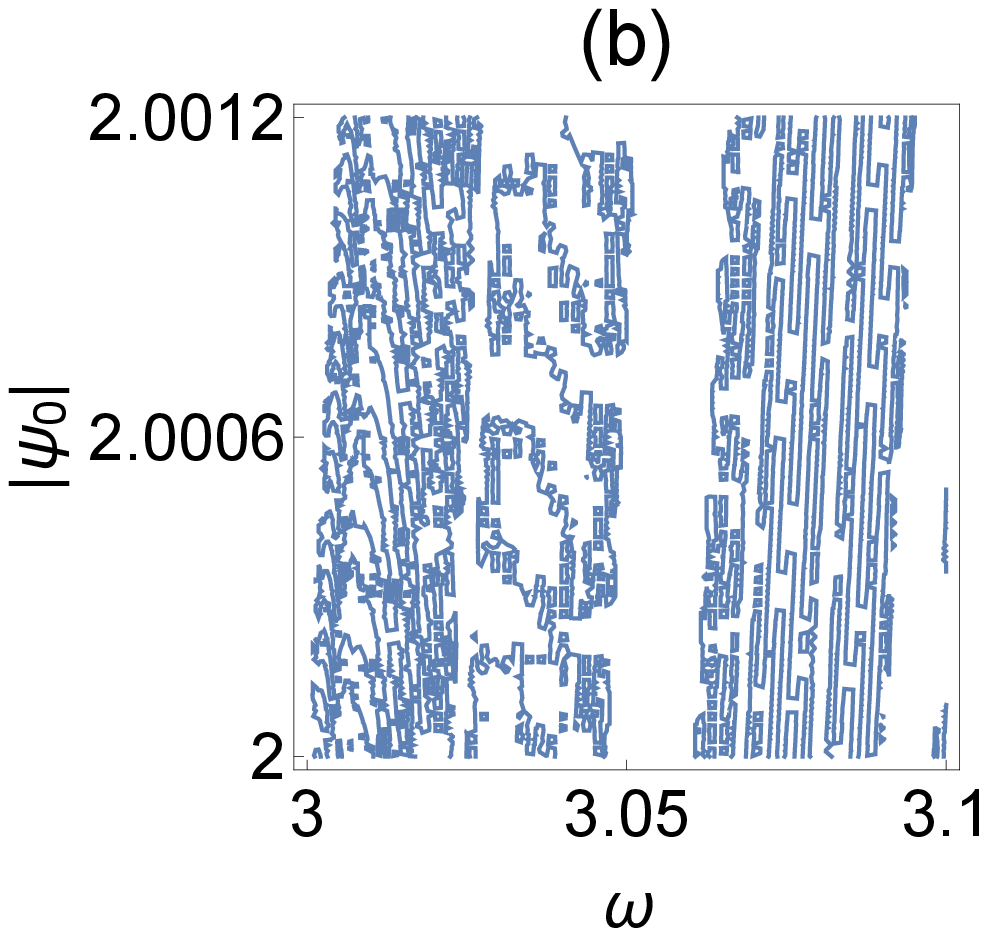}
\caption{ The plot (a) shows a fractal frequency spectrum when there are $7$ lattice sites at $\ds{g=1}$. The plot (b) shows the spectrum in a very narrower region depicted in the small red rectangle in the bottom left corner of the plot (a). We see a similar structure if we zoom in the plot. The fractal band is not continuous and every zoom is similar to the global plot.}
\end{figure}
The fractal band contains only a portion of the allowed frequencies. In fact, the system can have both $\it{continuum}$ and $\ds{fractal}$ band structures. Furthermore, the system can also have discrete frequency values at a given $|\psi_0|^2$ (well separated from these two bands). In the continuum band, the adjacent frequencies become so close together that they can be considered as a continuum. It is interesting to see that the solutions in the continuum band exhibits localization at the left edge while the solutions in the fractal band may or may not exhibit localization. A question arises. What are the upper and lower bounds of the continuum band? Let us answer this question firstly for the real-valued frequencies $\ds{\omega_R}$. To get the answer, we look for solutions by assuming that $\ds{   \left|   \psi_j    \right|  }  $ decreases in magnitude monotonically, $\ds{     \left| \psi_{j+1}     \right|   <  \left| \psi_{j}     \right|  }$, which implies that the OBC is satisfied since the field amplitude asymptotically approaches zero. This assumption is satisfied if $\ds{ 0<  \left|   \frac{\psi_{j+1}}{\psi_j}   \right|=    \left|   \omega_R-g|\psi_j|^2   \right|  <1}$ for every $\ds{j \geq 0}$. One can readily find that $\ds{\omega_R<1}$ and $\ds{\omega_R-g |  \psi_{0}|^2> - 1 }$. Since $N\rightarrow \infty  $, the frequency can take continuous values between them. We also add the point $\ds{\omega_R=1}$, at which there exists a localized solution according to Eq. (\ref{ddekcy23uds2}). 
\begin{equation}\label{77c57istsbcdu} 
 -(1-g |  \psi_{0}|^2 ) <  \omega_R\leq1 
\end{equation}
This continuum band lies below the fractal band since $\ds{ \omega_{+}>2.6 }$. Unlike the fractal band, the continuum band disappears at the critical value $\ds{g |  \psi_0|^2=2  }$. We note that continuum bands exist also for the linear system, but the fractal band is unique to the nonlinear one. Since $\ds{   \left|   \psi_j    \right|  }  $ are monotonically decreasing, the continuum band has only localized modes at the left edge. However, no such localized solutions are available for PBC. This nonlocal effect of the open edge is a signature of NHSE. There are two main differences between the linear and nonlinear NHSE. Firstly, all eigenstates are localized at the left edge in the linear case, $\ds{g=0}$, (since no fractal band appears) while all stationary solutions in the continuum band are localized at the left edge in the nonlinear case. Secondly, the nonlinear NHSE disappears when  $\ds{g |  \psi_0|^2\geq2  }$. This is already expected since the nonlinear interaction can be considered as an effective site-dependent potential, which can destroy NHSE beyond a critical strength. \\
We also numerically check that the formula (\ref{77c57istsbcdu}) is correct (the field amplitude diverges if the frequency value is chosen just below or above the lower or upper bounds, respectively). As a result, one can continuously get stationary solutions $\psi_j$ in the continuum band by either varying $\ds{\omega_R}$ at fixed $\psi_0$ according to (\ref{77c57istsbcdu}) or varying $\psi_0$ at fixed $\ds{\omega_R}$ according to $\ds{ 0<g| \psi_0|^2 <1+\omega_R}$. Note also that $g|\psi_0|^2=1$ are fixed at $\omega_R=1 $.  \\
Let us next find the boundaries of the imaginary part of the complex continuum band at fixed $\ds{\omega_R}$. We generalize the above condition to $\ds{   \left|   \frac{\psi_{j+1}}{\psi_j}   \right|=    \left|   \omega_R +i~\omega_I-g|\psi_j|^2  \right|  <1}$. It is satisfied if $\ds{ |\omega|=\sqrt{  \omega_R^2+\omega_I^2}<1  }$ and $\ds{ (\omega_R-g |  \psi_{0}|^2)^2  +\omega_I^2<1 }$. Therefore, the following two equations should be solved together to find the upper and lower bounds of $\ds{\omega_I}$ 
\begin{eqnarray}\label{c580dsldu} 
 \omega_I^2&<&1-   (\omega_R-g |  \psi_{0}|^2)^2  \nonumber\\
\omega_I^2&<&1- \omega_R^2
\end{eqnarray}
It is interesting to see that $\ds{\omega_I=0}$ at $\ds{\omega_R=1}$. In other words, the point $\ds{\omega_R=1}$ is the transition point from real to complex valued frequencies. Note also that $g|\psi_0|^2$ is fixed at $\ds{\omega_R=1}$ while it can be varied at other values of $\ds{\omega_R}$ in the continuum band (the same $g|\psi_0|^2$ but with a different $\omega_R$ still yields a solution, or vice versa). Recall that an exceptional point in a linear system determines the transition point from real to complex valued eigenvalues and there is a coalescing eigenstate at the exceptional point. Therefore, one may consider the point at $\ds{\omega_R=1}$ as a kind of nonlinear exceptional point.\\ 
Having studied the fully nonreciprocal case, consider now the case with $\ds{\gamma\neq0}$. In this case, obtaining whole family of solutions analytically is almost impossible. Simple solutions such as (\ref{ddekcy23uds2}, \ref{cs3aek2}) are absent. But fractal and continuum bands can still appear. We are interested in continuum bands as they have localized stable solutions. To find its upper and lower values, consider the ratio $\ds{   \left|   \frac{\psi_{j+1}}{\psi_j}   \right|= \left|   \omega_R-g|\psi_j|^2- \gamma  \frac{\psi_{j-1}}{\psi_j}       \right|   }$ for real valued frequencies. Assume that $\ds{   \left|   \psi_{j+1}    \right|   < \left|   \psi_{j}    \right|    }  $ for all $j$. This condition is satisfied if $\ds{     -(1+\gamma -g |  \psi_{0}|^2 )<\omega_R\leq 1+\gamma}$.  Note that solutions with $\ds{ \omega_R>1+\gamma  }$ either repeat themselves at large values of $j$ or blow up. Unfortunatelly, the above assumption is incomplete as there exists other solutions where $\ds{ | \psi_j|   }$ is not maximum at the left edge. Instead, $\ds{|\psi_j|}$ increases for a few sites and then monotonically decreases: $\ds{   \left|   \psi_{j+1}    \right|   < \left|   \psi_{j}    \right|      }  $ for all $\ds{j>n }$, where $\ds{n}$ depends on $\ds{\gamma}$ and can be found numerically. Then
\begin{equation}\label{ayr8940cu} 
 -(1+\gamma-g |  \psi_{n}|^2 )<\omega_R\leq1 +\gamma
\end{equation}
We numerically check that this formula derived based on the above assumptions is valid as long as $\ds{ \gamma  }$ is small. We further numerically find that $n<10$ for small values of $\ds{\gamma}$. As $\ds{ \gamma  }$ approaches $\ds{1}$, the deviation from this formula is significant. Additionally, the continuum band is broken into some other bands with forbidden band gap. For example, at $\ds{\gamma=0.8}$ and $\ds{ g=\psi_0=1}$, the forbidden band occurs when $\ds{0.60<\omega_R<0.82}$. Recall that even for the linear system $g=0$, NHSE occurs for highly nonreciprocal lattice and hence our analysis for small values of $\ds{\gamma}$ is good enough to study nonlinear NHSE. Let us now briefly discuss the upper and lower bound for $\ds{\omega_I}$ since it is challenging to obtain them analytically. They strongly depend on $\ds{\omega_R}$, $\ds{\gamma}$ and $\ds{|\psi_0|}$. One can easily say that the imaginary part of the continuum band shrinks more and more with increasing $\ds{\gamma}$ and $\ds{\omega_I}$ become zero at the Hermitian limit $\ds{\gamma=1}$. We numerically see that $\ds{|\omega_I|< 1-\gamma }$ for small values of $\ds{  \omega_R  }$. 

\section{ A non-Hermitian Ablowitz-Ladik model}

Let us now study nonlinear NHSE for another model. Consider the following non-Hermitian extension of the Ablowitz-Ladik (AL) equation \cite{ALmodel}  for a further understanding of NHSE in nonlinear domain
\begin{equation}\label{rof64oalk2} 
(1+g ~ |\psi_{j}  |^2) ~(\psi_{j+1}+\gamma ~\psi_{j-1}  )=\omega ~\psi_{j} 
\end{equation} 
where $\ds{  \psi_j }$ is the complex field amplitude at the lattice site $\ds{ j=0,1,...,N }$,  the parameter $\ds{ 0\leq \gamma<1 }$ is the non-Hermitian degree, $\ds{ g >0  }$ is the positive nonlinear interaction strength and $\ds{  \omega }$ is the frequency.\\
In the case of PBC, the delocalized solutions of the form $\ds{ \psi_j=  \psi_0 ~e^{ik j} }$ satisfy Eq. (\ref{rof64oalk2}). The corresponding frequencies read $\ds{ \omega_k=(1+g|\psi_0|^2)(  e^{ik } +\gamma~e^{ik }   ) }$. In the case of OBC, $\ds{\psi_{-1}=\psi_{N+1}=0}$, we expect the frequency spectra to be drastically changed and the corresponding solutions to be localized at the edge due to NHSE. We can check them simply at a particular value $\ds{\gamma=0}$. In this case, we recursively solve the nonlinear recurrence relation $\ds{    \psi_{j+1}=\frac{\omega~\psi_j}{1+g  |\psi_{j}  |^2 }   }$. We set the initial value, $\ds{  \psi_{0} }$, at the left edge and obtain the other terms of the sequence from the preceding terms. It is interesting to see that the OBC at the right edge, $\ds{\psi_{N+1}=0}$, is exactly satisfied only when $\ds{ \omega=0}$. In other words, there exists only one solution, which is localized perfectly at the left edge and given by $\ds{ \psi_j=  \psi_0 ~ \delta_{0,j}}$. This means that a {\it {nonlinear exceptional point}} of order (N+1) appears for OBC as all nonlinear stationary solutions coalesce. As a result, we say that NHSE can occur for this model since the systems with PBC and OBC have completely different spectra at $\ds{\gamma=0}$ and the former one has extended solutions while the latter one has a unique localized solution. \\
One can intuitively expect that NHSE can still be visible for small values of $\ds{\gamma}$. However, no exceptional point occurs when $\ds{ \gamma\neq0}$ and the total number of solutions grow exponentially with $\ds{N}$. Finding the frequency spectrum numerically for a long open lattice is challenging, let alone analytical calculations. Instead of finding the spectrum exactly for a finite lattice from scratch, we study the limiting case $\ds{  N  \rightarrow\infty}$. In Hermitian systems, mathematically extending the boundary of a very long lattice to infinity just changes the spectrum perturbatively \cite{aciklama}. We have recently shown that this may not be the case in linear non-Hermitian systems \cite{csljax3}. In other words, an infinitely long and a very long lattices can have dramatically different spectrum. Below we will show that this is also true for our non-linear model. 
\\$\ds{ \it{Semi-infinite ~AL ~lattice :}  }$ Consider a semi-infinite lattice with $\ds{\psi_{-1}= \psi_{\infty} = 0  } $. The nonlinear recurrence relation reads $\ds{    \psi_{j+1}=\frac{\omega~\psi_j      }{1+g  |\psi_{j}  |^2  }  -\gamma~\psi_{j-1}   }$. For a given $\ds{\psi_0}$, the successive terms such as $\ds{  \psi_1=\frac{\omega~\psi_0     }{1+g  |\psi_{0}  |^2  }    }$ and $\ds{  \psi_2=\frac{\omega^2~(1+g  |\psi_{0}  |^2 )~\psi_0     }{ 1+2 g  |\psi_{0}  |^2    +g \omega^2   |\psi_{0}  |^2 +g^2   |\psi_{0}  |^4   }  -\gamma~\psi_{0}    }$ can be obtained. Here, we look for solutions satisfying the monotonicity condition of $\ds{|\psi_j|}$: $\ds{ | \psi_{j+1}|<|\psi_j|}$ for all $j\geq0$. In this way, the condition $\ds{\psi_{\infty} = 0  } $ can be satisfied. Suppose first that the frequencies are real valued, $\ds{\omega=\omega_R}$. The monotonicity condition is satisfied if $\ds{\omega_R}$ takes continuous values according to
\begin{equation}\label{idjhv5647c} 
-1-\gamma<\omega_R< 1+\gamma
\end{equation} 
This represents a continuum band and the solutions are localized at the left edge since $\ds{|\psi_j|}$ monotonically decreases. This implies that NHSE occurs in this continuum band.\\
It is interesting to study this expression at the particular value $\ds{\gamma=0}$. In this case, Eq. (\ref{idjhv5647c}) yields $\ds{|  \omega |<1}$. This seems to contradict to our earlier result that an exceptional point with zero frequency $\ds{\omega=0}$ occurs at $\ds{\gamma=0}$. In fact, there is no ambiguity since non-Hermitian systems can have drastically different spectrum for finite and infinite lattices. The reason for this is as follows: The exceptional point, which occurs in a finite lattice, disappears in the semi-infinite lattice. As a result, there is only one coalescing solution for a finite lattice, no matter how long the lattice is, while there are infinitely many solutions if the lattice extends to infinity. Fortunately, the solutions for the infinitely long lattice can still be utilized as quasi-stationary solutions for a long finite lattice \cite{csljax3}. Quasi-stationary solutions are not exact stationary solutions but approximately satisfy OBC and stay almost stationary for a long time in a linear system. Here we show that such solutions can also be seen in nonlinear domain since the right open boundary condition is almost satisfied ($\ds{ |\psi_{N+1}| \approx 0  } $) in a long lattice when $\ds{|\omega|<1}$ For example, suppose $\ds{g=\psi_0=1}$, ${\gamma=0}$ and $\ds{N=100}$. Then we numerically find $\ds{ |\psi_{N+1}|  \approx 10^{-11}  } $ at $\ds{\omega=\mp0.8}$, $\ds{ |\psi_{N+1}| \approx 10^{-31}  } $ at $\ds{\omega=\mp0.5}$, $\ds{ |\psi_{N+1}| \approx 10^{-71}  } $ at $\ds{\omega=\mp0.2}$. These are very small numerical values and hence corresponding solutions can be considered as quasi-stationary solutions. Note that such solutions at $\gamma=0$ can still appear even when $\ds{\omega}$ are complex valued. In other words the right boundary condition is still satisfied if we replace $\ds{\omega\rightarrow \omega ~e^{i\theta}}$, where $\ds{0<\theta<2\pi}$ is an arbitrary number. One can generalize this concept to obtain quasi stationary states for small values of $\gamma$.  

\section{Conclusion}

We consider two specific nonlinear equations to study NHSE in nonlinear domain and show that open boundaries can have nonlocal effects on the nonlinear spectrum. We see that the type of nonlinear interaction determines how NHSE arises in long nonlinear lattices. For the non-Hermitian Ablowitz-Ladik lattice, the nonlinear exceptional point occurs and the system has only one localized solution when the open system is fully nonreciprocal ($\ds{\gamma=0}$). We further show that the solutions are still localized at the left edge when the open system is highly nonreciprocal. However, no nonlinear exceptional point occurs for periodical boundary conditions and the corresponding extended modes have complex valued spectra. We discuss that lattice size matters since the nonlinear exceptional point disappears for the semi-infinite lattice. This implies that infinitely and sufficiently long lattices can have drastically different spectrum. For the nonreciprocal discrete nonlinear Schrodinger lattice, there are two main differences between the linear and nonlinear NHSE. Firstly, all eigenstates are localized at the left edge in the linear case, $\ds{g=0}$, (since no fractal band appears) while all stationary solutions in the continuum band are localized at the left edge in the nonlinear case. Secondly, the nonlinear NHSE disappears for strong nonlinear interaction strength even if the system is fully nonreciprocal. This is because of the fact that the continuum band shrinks as the nonlinear interaction is increased and disappears if it is beyond a critical strength. There is another difference between linear and nonlinear NHSE. In linear systems, NHSE leads to funneling effect since every signal travels towards the boundary no matter where the lattice is excited. However, no funneling occurs in our nonlinear model due to the fractal band, which has unstable modes.


\begin{thebibliography}{0}
\bibitem{aciklama} This also true for continuous systems, for example, the Schrodinger equation is solved for the standard harmonic oscillator or hydrogen atom problems by assuming that the wave packet vanishes at infinity.
\end{thebibliography}
\end{document}